# A Review of Control Algorithms for Autonomous Quadrotors

**Andrew Zulu*, Samuel John**

Department of Mechanical and Marine Engineering, Polytechnic of Namibia, Windhoek, Namibia
Email: *azulu@polytechnic.edu.na, sjohn@polytechnic.edu.na

## Abstract

**The quadrotor unmanned aerial vehicle is a great platform for control systems research as its nonlinear nature and under-actuated configuration make it ideal to synthesize and analyze control algorithms. After a brief explanation of the system, several algorithms have been analyzed including their advantages and disadvantages: PID, Linear Quadratic Regulator (LQR), Sliding mode, Backstepping, Feedback linearization, Adaptive, Robust, Optimal, $L_1$, $H_\infty$, Fuzzy logic and Artificial neutral networks. The conclusion of this work is a proposal of hybrid systems to be considered as they combine advantages from more than one control philosophy.**

## 1. Introduction

The quadrotor unmanned aerial vehicle (UAV) are helicopters with four rotors typically designed in a cross configuration with two pairs of opposite rotors rotating clockwise and the other rotor pair rotating counter-clockwise to balance the torque. The roll, pitch, yaw and up-thrust actions are controlled by changing the thrusts of the rotors using pulse width modulation (PWM) to give the desired output.

Civilian use of unmanned/micro aerial vehicles (UAVs/MAVs) include aerial photography for mapping and news coverage, inspection of power lines, atmospheric analysis for weather forecasts, traffic monitoring in urban areas, crop monitoring and spraying, border patrol, surveillance for illegal imports and exports, fire detection and control, search and rescue operations for missing persons and natural disasters.

In this work, the focus is to realize the best configuration and control scheme of the quadrotor MAV for ap-plication in game counting in the protected game reserves in Africa. The current situation in most African countries is that researchers and game rangers either count from road vehicles by driving by the road or use conventional helicopters. The former is ineffective as the count is limited to only animals coming close to the road whereas the later leads to inaccurate counts as the animals run for safety when a helicopter hovers above them. Several advantages accrue to the quadrotor which has led to much attention from researchers. This is mainly due to its simple design, high agility and maneuverability, relatively better payload, vertical take-off and landing (VTOL) ability.

*Corresponding author.

The quadrotor does not have complex mechanical control linkages due to the fact that it relies on fixed pitch rotors and uses the variation in motor speed for vehicle control [1]. However, these advantages come at a price as controlling a quadrotor is not easy because of the coupled dynamics and its commonly under-actuated design configuration [2]. In addition, the dynamics of the quadrotor are highly non-linear and several uncertainties are encountered during its missions [3], thereby making its flight control a challenging venture. This has led to several control algorithms proposed in the literature. In this work, a review of the prominent controllers applied to the quadrotor is reviewed.

The rest of this paper is organized as follows: Section 2 provides a brief description of the mathematical model of the quadrotor; Section 3 gives a review of the popular controllers proposed for the quadrotor and a discussion of hybrid control systems; in Section 4 a tabular comparison is presented and the conclusions are given in Section 5.

## 2. Mathematical Model

In this section a brief description of the main mathematical equation for the quadrotor is explained.

In flight mode, the quadrotor is able to generate its lift forces by controlling the fixed pitch angle rotor speeds. **Figure 1** presents the schematic diagram. The center of mass is the origin O of the coordinate system (see **Figure 1**) and the forward direction is arbitrarily fixed in the *x*-axis. To lift off from ground, all four rotors are rotated at the same speed in the sense shown. A total thrust equal to the weight of the system stabilizes the system in hover. To roll about the *x*-axis, differential thrust of rotors 2 and 4 is applied. To pitch about the *y*-axis, differential thrust of rotors 1 and 3 is applied. The yaw motion is achieved by the differential thrusts of the opposite rotors (1/3 or 2/4) while also adjusting the constant thrusts of the remaining opposite pairs to maintain altitude.

Using the Newton-Euler formalism, the equations of motion of the quadrotor are given in [4]:

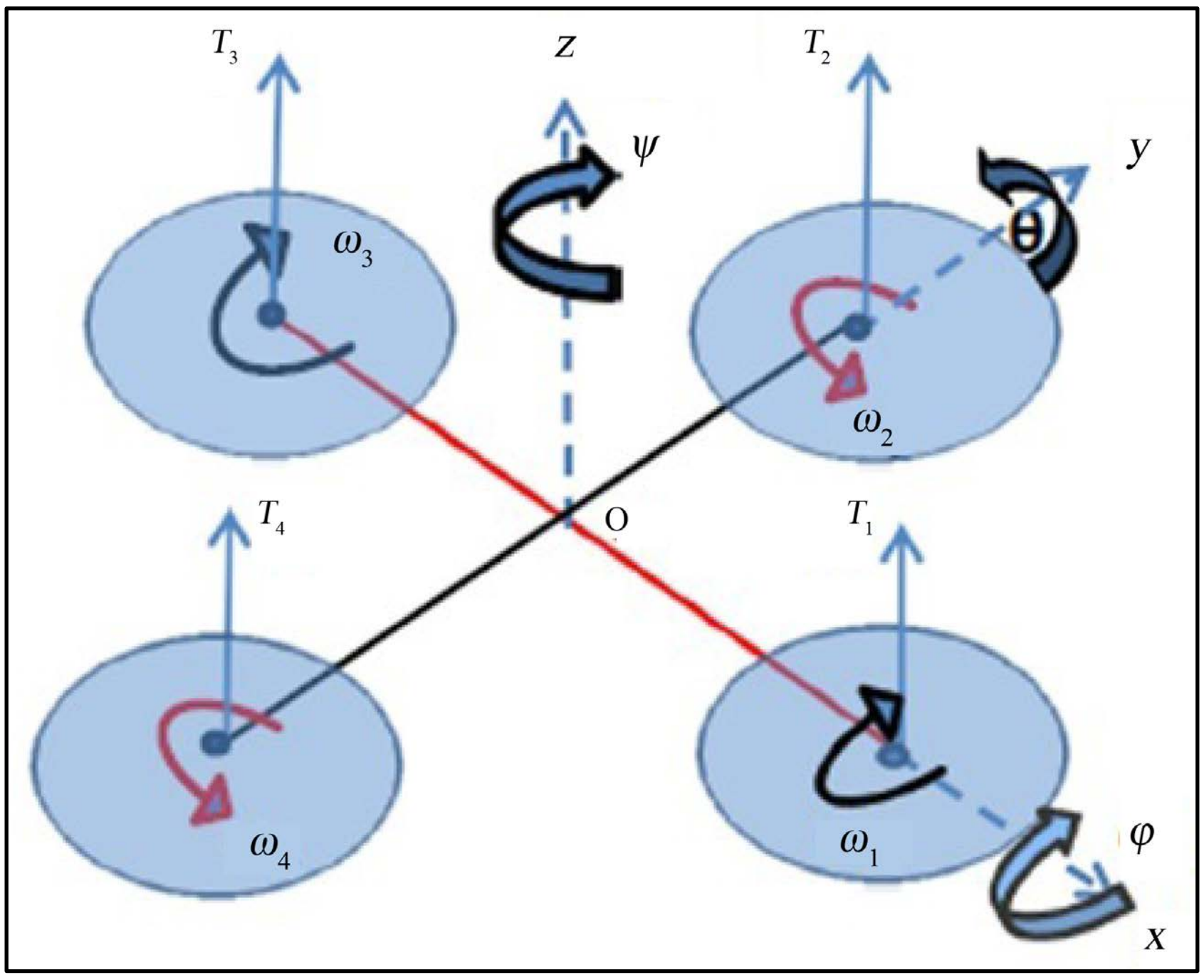

**Figure 1.** Schematic of quadrotor showing rotor configurations.

$$\ddot{\phi} = \dot{\theta}\dot{\psi}\left(\frac{I_y - I_z}{I_x}\right) - \frac{J_r}{I_x}\dot{\theta}\Omega + \frac{l}{I_x}U_2,$$
$$\ddot{\theta} = \dot{\phi}\dot{\psi}\left(\frac{I_z - I_x}{I_y}\right) - \frac{J_r}{I_y}\dot{\phi}\Omega + \frac{l}{I_y}U_3,$$
$$\ddot{\psi} = \dot{\phi}\dot{\theta}\left(\frac{I_x - I_y}{I_z}\right) + \frac{1}{I_z}U_4,$$
$$\ddot{z} = -g + (\cos\phi\cos\theta)U_1/m,$$
$$\ddot{x} = (\cos\phi\sin\theta\cos\psi + \sin\phi\sin\psi)U_1/m,$$
$$\ddot{y} = (\cos\phi\sin\theta\sin\psi - \sin\phi\cos\psi)U_1/m.$$

where $\Phi$, $\theta$ and $\psi$ are the roll, pitch and yaw angles respectively; $I_x$, $I_y$, $I_z$ are the mass moments of inertia in the $x$, $y$, and $z$ axes respectively; $J_r$ is the rotor inertia; $\Omega$ is the angular velocity of the rotor; $l$ is the length of the rotor arm from the origin of the coordinate system; and the inputs to the four rotors are given by the expressions:

$$U_1 = b\left(\Omega_1^2 + \Omega_2^2 + \Omega_3^2 + \Omega_4^2\right),$$
$$U_2 = b\left(-\Omega_2^2 + \Omega_4^2\right),$$
$$U_3 = b\left(\Omega_1^2 - \Omega_3^2\right),$$
$$U_4 = d\left(-\Omega_1^2 + \Omega_2^2 - \Omega_3^2 + \Omega_4^2\right).$$

where $b$ and $d$ are thrust and drag coefficients respectively. For detailed derivation of the equations, consult [4].

## 3. Survey of Control Algorithms

Due to the nature of the dynamics of the quadrotor, several control algorithms have been applied to it. As to be expected, each control scheme has its advantages and disadvantages. The control schemes used could be broadly categorized as linear and non-linear control schemes. In this review a broad range of controllers within these categories are discussed.

### 3.1. Proportional Integral Derivative (PID)

The PID controller has been applied to a broad range of controller applications. It is indeed the most applied controller in industry [5]. The classical PID linear controller has the advantage that parameter gains are easy to adjust, is simple to design and has good robustness. However some of the major challenges with the quadrotor include the non-linearity associated with the mathematical model and the imprecise nature of the model due to unmodeled or inaccurate mathematical modeling of some of the dynamics. Therefore applying PID controller to the quadrotor limits its performance.

A PID controller was used for the attitude control of a quadrotor in [6], while a dynamic surface control (DSC) was used for the altitude control. Applying Lyapunov stability criteria, Lee *et al*. were able to prove that all signals of the quadrotor were uniformly ultimately bounded. This meant that the quadrotor was stable for hovering conditions. From the simulation and experimental plots however, it reveals the PID controller to have performed better in the pitch angle tracking, whereas large steady state errors could be observed in the roll angle tracking.

In another work by Li and Li [7], a PID controller was applied to regulate both position and orientation of a quadrotor. The PID parameter gains were chosen intuitively. The performance of the PID controller indicated relatively good attitude stabilization. The response time was good, with almost zero steady state error and with a slight overshoot.

It is generally established in the literature that the PID controller has been successfully applied to the quadrotor though with some limitations. The tuning of the PID controller could pose some challenges as this must be con-ducted around the equilibrium point, which is the hover point, to give good performance.

**Figure 2** shows the general block diagram of a PID controller for the quadrotor.

## 3.2. Linear Quadratic Regulator/Gaussian-LQR/G

The LQR optimal control algorithm operates a dynamic system by minimizing a suitable cost function. Boubdallar and co-researchers applied the LQR algorithm to a quadrotor and compared its performance to that of the PID controller in [8]. However, the PID is applied on the quadrotor simplified dynamics and the LQR on the complete model. Both approaches provided average results but it implicitly was clear that the LQR approach had better performance considering the fact that it was applied to a more complete dynamic model.

A simple path-following LQR controller was applied by Cowling *et al.* in [9] on a full dynamic model of the quadrotor. It was shown that accurate path following was achieved in simulation using optimal real-time trajectories despite the presence of wind and other disturbances. The controller seemed to lose tracking after avoiding an obstacle. Its performance in the presence of many obstacles still needed to be analyzed.

In combination with a Linear Quadratic Estimator (LQE) and Kalman Filter, the LQR algorithm transforms into the Linear Quadratic Gaussian (LQG). This algorithm is for systems with Gaussian noise and incomplete state information. The LQG with integral action was applied in [10] for stabilization of attitude of a quadrotor with good results in hover mode. The advantage of this LQG controller is the fact that you do not need to have complete state information to implement it.

**Figure 3** shows the general block diagram of an LQG controller for the quadrotor.

## 3.3. Sliding Mode Control (SMC)

Sliding mode control is a nonlinear control algorithm that works by applying a discontinuous control signal to the system to command it to slide along a prescribed path. Its main advantage is that it does not simplify the dy-namics through linearization and has good tracking. The sliding mode controller was applied to stabilize cascaded under-actuated systems in [11]. The quadrotor system was subdivided into the full-actuated and under-actuated systems. The under-actuated system, to which SMC was applied, was further subdivided into under-actuated subsystems. Results showed good stability and robustness of the system. Chattering effect of SMC was observed but minimized with a continuous approximation of a pre-determined "sign" function.

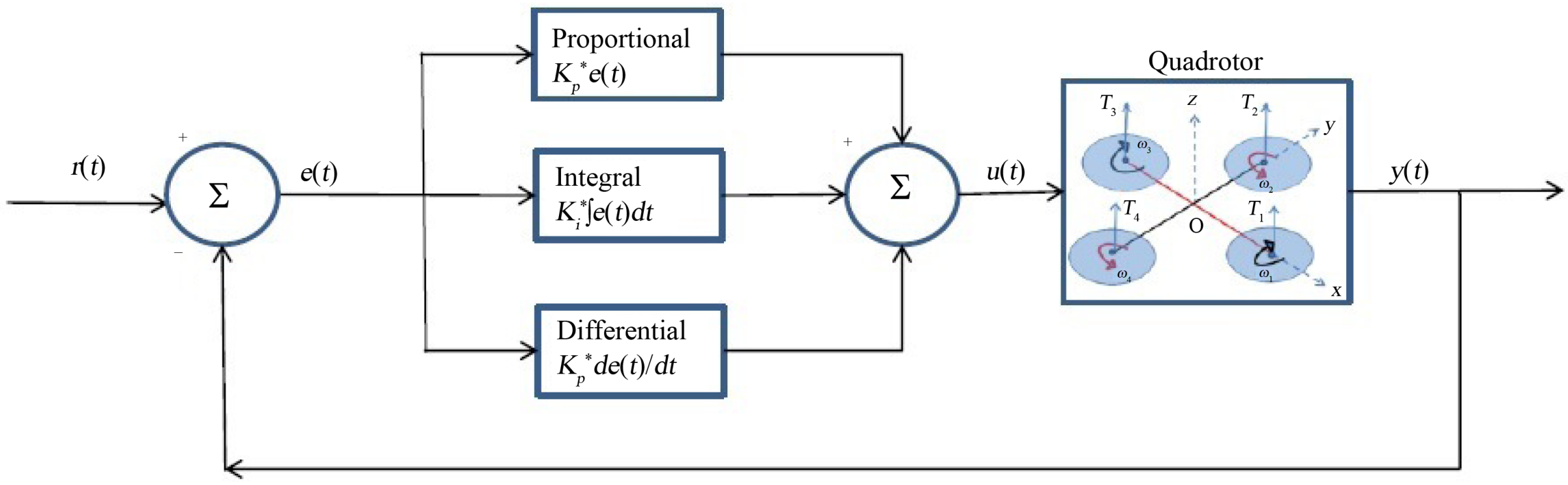

**Figure 2.** Block diagram of PID controller applied to the quadrotor.

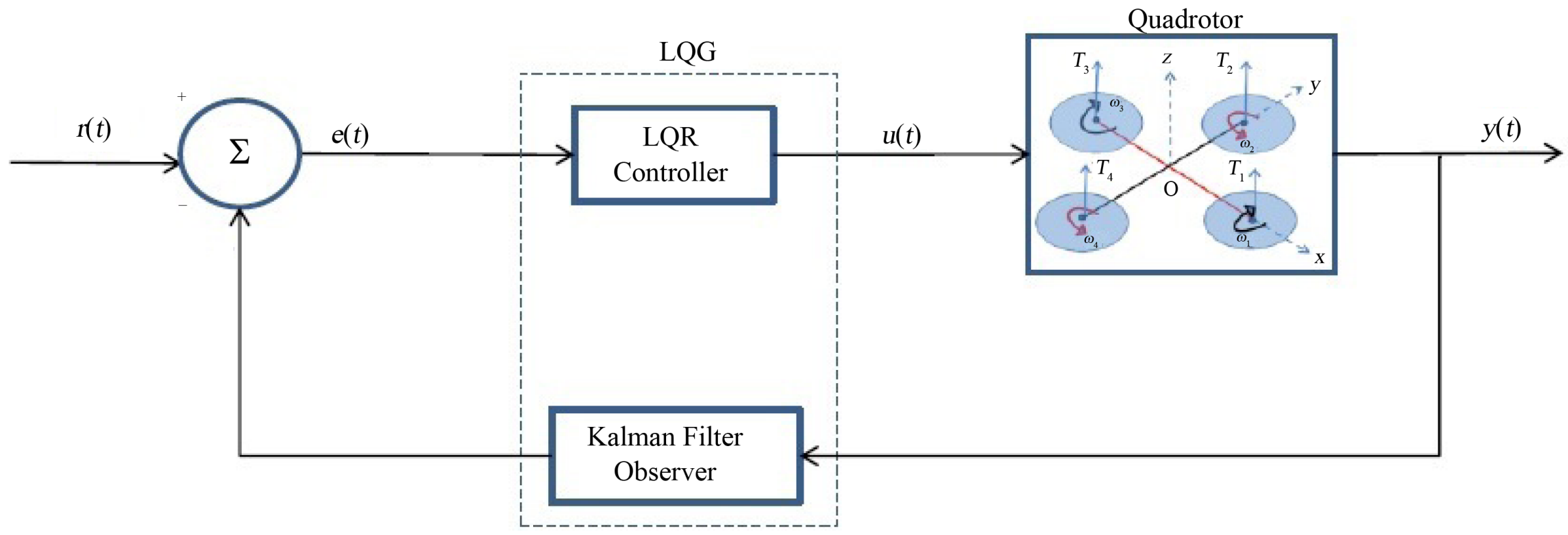

**Figure 3.** Block diagram of LQG controller applied to the quadrotor.

A sliding mode controller based on Lyapunov stability theory is articulated by Runcharoon and Srichatrapimuk in [12]. The SMC controller was able to stably drive the quadrotor to a desired position and yaw. Tracking was equally good with injected noise, which showed good robustness for the SMC controller.

**Figure 4** shows the general block diagram of an SMC controller for the quadrotor.

### 3.4. (Integrator) Backstepping Control

Backstepping control is a recursive algorithm that breaks down the controller into steps and progressively stabilizes each subsystem. Its advantage is that the algorithm converges fast leading to less computational resources and it can handle disturbances well. The main limitation with the algorithm is its robustness is not good. Madani and co-researchers [13] applied backstepping control to stabilize a quadrotor system consisting of an under-actuated, fully actuated and propeller subsystems. Good tracking was achieved for position and yaw angle. Using Lyapunov stability theory, roll and pitch angles were stabilized.

The backstepping approach was also applied in [1] for attitude stabilization of a quadrotor. Using Lyapunov stability analysis, the closed-loop attitude system was found to be asymptotically stable with all states uniformly ultimately bounded in the presence of external disturbance. It was also implicit that the quaternion formulation also helped in the computational side for stabilization in addition to avoiding singularity.

To increase robustness (to external disturbances) of the general backstepping algorithm, an integrator is added and the algorithm becomes Integrator backstepping control as articulated by Fang and Gao in [14]. The integral approach was shown to eliminate the steady-state errors of the system, reduce response time and restrain overshoot of the control parameters.

### 3.5. Adaptive Control Algorithms

Adaptive control algorithms are aimed at adapting to parameter changes in the system. The parameters are either uncertain or varying with time. A continuous time-varying adaptive controller, which shows good performance, was implemented by Diao *et al*. in [15] with known uncertainties in mass, moments of inertia and aerodynamic damping coefficients.

In other work by Palunko *et al*., an adaptive control scheme using feedback linearization (FBL) is implemented [16] for quadrotors with dynamic changes in the center of gravity. It was observed that when the center of gravity changes, PD and regular feedback linearization techniques were not able to stabilize the system but the adaptive controller was able to stabilize it. The paper touches on a very important futuristic quadrotor that will self-configure in real time when the center of gravity changes.

An adaptive control technique based on rectilinear distance $(L_1)$ norm was used in [17] with a tradeoff between control performance and robustness. The modified (linearized) model was able to compensate for constant and moderate wind gusts.

**Figure 5** shows the general block diagram of an adaptive controller for the quadrotor clearing showing the parameter estimator and quadrotor model.

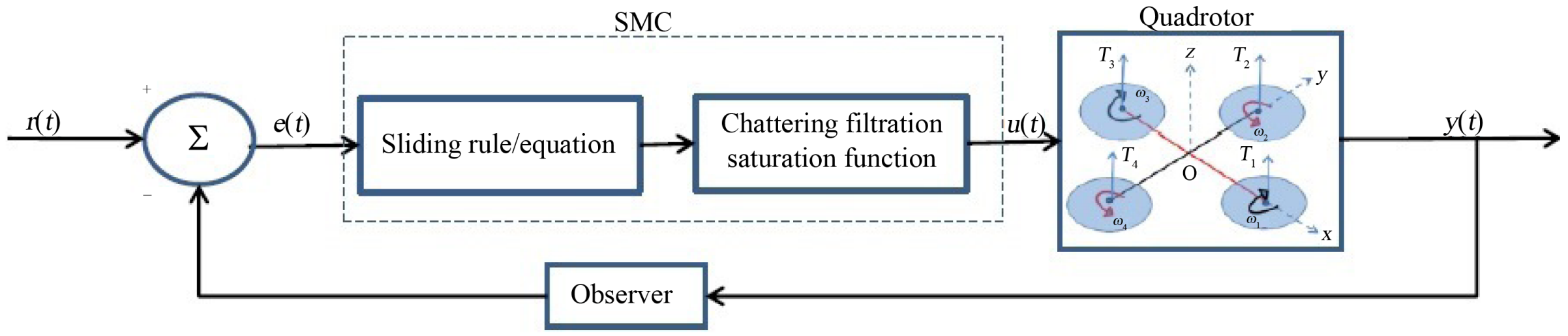

**Figure 4.** Block diagram of an SMC controller applied to the quadrotor.

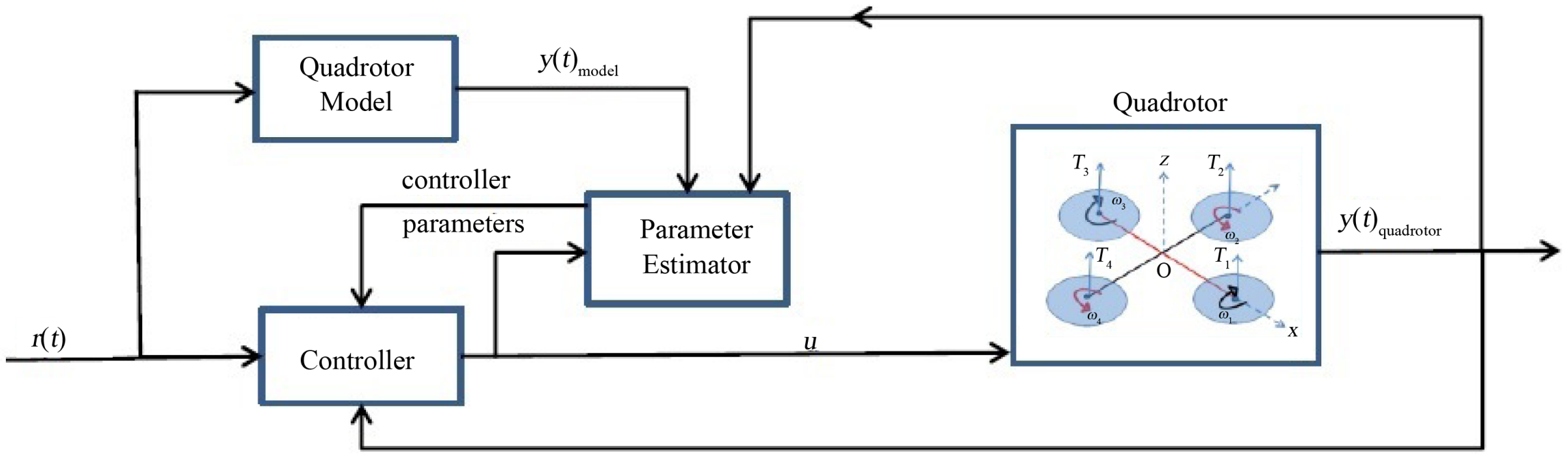

**Figure 5.** Block diagram of an adaptive controller applied to the quadrotor.

## 3.6. Robust Control Algorithms

Robust control algorithms are designed to deal with uncertainty in the system parameters or disturbances. This guarantees controller performance within acceptable disturbance ranges or un-modeled system parameters. A major limitation normally observed with robust controllers is poor tracking ability. A robust controller was implemented by Bai *et al*. in [18] for an attitude controller subsystem based on linear control and robust compensation. Experimental results validated the effectiveness of the controller for controlling attitude.

In another work by Tony and Mackunisy a robust tracking algorithm is presented [19] and achieves asymptotic stability in the presence of parametric uncertainties and unknown nonlinear disturbances. This was achieved with inexpensive control law, no observers, functional approximators or online adaptive updating laws.

## 3.7. Optimal Control Algorithms

Optimization algorithms are designed to minimize a variable and get the best cost function from a set of alternatives. A specialized case of optimization is known as convex optimization. This is a mathematical optimization technique that deals with minimizing a convex variable in a convex set. The common optimal algorithms include LQR, $L_1$, $H_\infty$ and Kalman filter. A major limitation of optimization algorithms is generally their poor robustness. A controller that is both relatively robust and $L_1$-optimal was applied in [20] for tracking of both attitude and heading. The performance of the controller was experimentally validated and was efficient in error minimization and rejection of persistent disturbances.

An optimization algorithm based on $H_\infty$ looping was applied by Falkenberg *et al*. in [21] to a quadrotor with simplified dynamics for iterative parameter identification and attitude control. The algorithm had superior performance with respect to disturbance rejection even under disturbance saturation conditions. However, the ant-windup compensator based on Riccati equations did not seem to show computational efficiency.

An integral optimal predictive $H_\infty$ controller was implemented by Raffo *et al*. in [22] for stabilization of rotational movement of a quadrotor and for path following (using model predictive control). The controller achieved sustained disturbance rejection and good robustness. Integral action was key to achieving good tracking in this algorithm.

## 3.8. Feedback Linearization

Feedback linearization control algorithms transform a nonlinear system model into an equivalent linear system through a change of variables. Some limitation of feedback linearization is the loss of precision due to linearization and requiring an exact model for implementation. Output feedback linearization was implemented as an adaptive control strategy in [16] for stabilization and trajectory tracking on a quadrotor that had dynamic changes of its center of gravity. The controller was able to stabilize the quadrotor and reconfigure it in real time when the center of gravity changed.

Feedback linearization and input dynamic inversion were implemented by Roza and Maggiore in [23] to design a path-following controller which allowed the designer to specify the speed profile and yaw angle as a function of displacement along the path. Two simulation cases with the quadrotor travelling at different speeds along the path were considered. Both cases showed convergence of velocity and yaw angle.

Feedback linearization was compared with adaptive sliding mode control in [24]. The feedback controller, with simplified dynamics, was found to be very sensitive to sensor noise and not robust. The sliding mode controller performed well under noisy conditions and adaptation was able to estimate uncertainties such as ground effect.

Hence, feedback linearization nonlinear control shows good tracking but poor disturbance rejection. However, feed-back linearization applied with another algorithm with less sensitivity to noise give good performance.

### 3.9. Intelligent Control (Fuzzy Logic and Artificial Neural Networks)

Intelligent control algorithms apply several artificial intelligence approaches, some biologically-inspired, to control a system. Examples include fuzzy logic, neural networks, machine learning, and genetic algorithm. They typically involve considerable uncertainty and mathematical complexity. This complexity and abundant computational resources required are limitations to the use of intelligent systems.

Intelligent control is not limited to fuzzy logic and neural networks but the two are the most widely used. Many other algorithms exist and continue to be formulated. Fuzzy logic algorithms deal with many-valued logic, not discrete levels of truth. An intelligent fuzzy controller was applied by Santos *et al*. in [25] to control the position and orientation of a quadrotor with good response in simulation. However, a major limitation of this work was the trial and error approach for tuning of input variables.

Artificial neural networks are biologically inspired by the central nervous system and brain. A robust neural networks algorithm was applied to a quadrotor by Nicol *et al*. in [26] to stabilize against modeling error and considerable wind disturbance. The method showed improvements with respect to achieving desired attitude and reducing weight drift.

Output feedback control was implemented on a quadrotor using neural networks [27] for leader-follower quadrotor formation to learn the complete dynamics of the UAV including un-modeled dynamics. A virtual neural network control was implemented to control all six degrees of freedom from four control inputs. Using Lyapunov stability theory, it was shown that the position, orientation, velocity tracking errors, observer estimation errors and virtual control were semi-globally uniformly ultimately bounded.

An adaptive neural network scheme was applied in [28] for quadrotor stabilization in the presence of a sinusoidal disturbance. The proposed solution of two parallel single hidden layers proved fruitful as reduced tracking error and no weight drift were achieved.

**Figure 6** shows the general block diagram of a fuzzy logic controller implemented on the quadrotor.

### 3.10. Hybrid Control Algorithms

It is evident that even the best linear or nonlinear algorithms had limitations and no single controller had it all. Researchers have tackled this by combining the philosophies of one or more algorithms. Here are few examples, which are not in any way exhaustive of what is in literature.

A hybrid fuzzy controller with backstepping and sliding mode control was implemented in [29] and successfully eliminated chattering effect of the sliding mode control algorithm. A feedback linearization controller was applied parallel with a high-order sliding mode controller to a quadrotor in [30]. The sliding mode controller was used as an observer and estimator of external disturbances. The system showed good disturbance rejection and robustness.

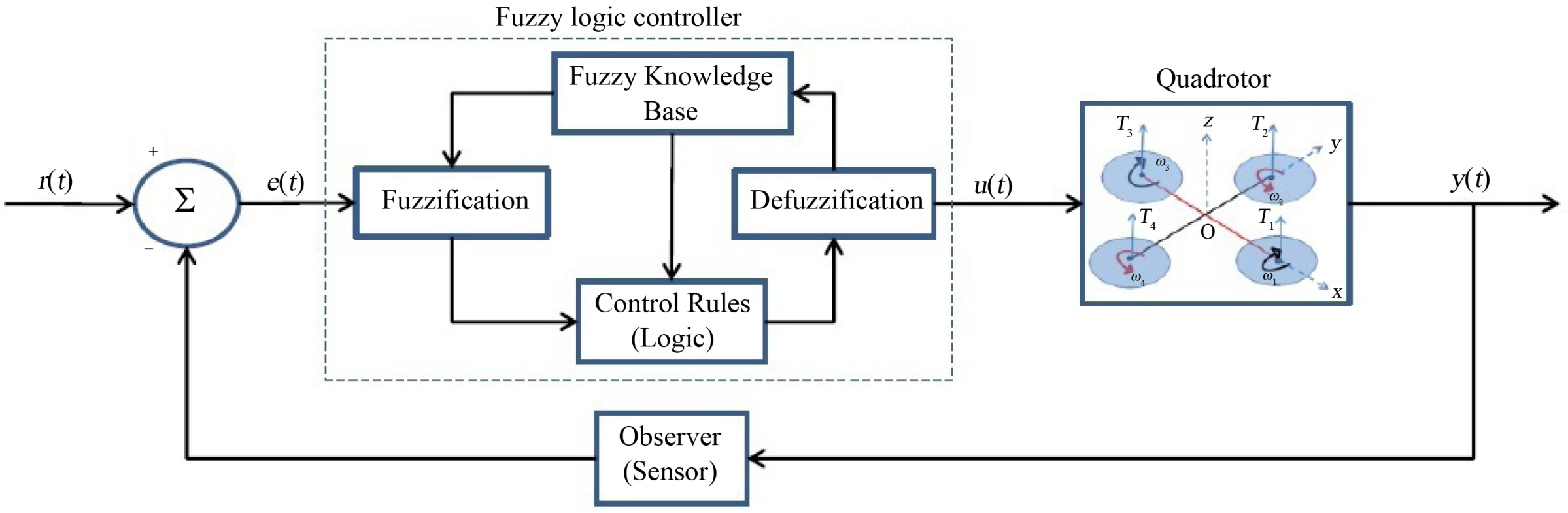

**Figure 6.** Block diagram an FLC controller applied to the quadrotor.

Adaptive control via backstepping combined with neural networks was presented by Madani and co-researchers in [31]. The backstepping was used to achieve good tracking of desired translational positions and yaw angle whilst maintaining stability of roll and pitch angles. Neural networks were used to compensate for unmodeled dynamics. The major contribution of the paper was the fact that the controller did not require the dynamic model and other parameters. This meant greater versatility and robustness.

## 4. Comparison of Control Algorithms

**Table 1** summarizes the comparison of the various algorithms as applied to quadrotors with all things being equal. The performance of a particular algorithm depends on many factors that may not even be modeled. Hence, this table serves as "fuzzy" guide in accordance with what is presented in this paper and common knowledge.

## 5. Discussion and Conclusion

This survey acts as a stepping stone into further work that will be conducted by the authors on control of quadrotors with tilting propellers. The tilting propellers further increase the number of actuated inputs to give six control inputs for six control outputs, which makes the quadrotor fully actuated. This research direction is relatively new with few examples in literature but with very promising performance achievement (see [32]-[34]). This work will focus on optimal control algorithms to achieve asymptotic stability of the quadrotor under any conditions.

**Table 1.** Comparison of quadrotor control algorithms.

| Control Algorithm | Characteristic | | | | | | | | | | | | |
|---|---|---|---|---|---|---|---|---|---|---|---|---|---|
| | Robust | Adaptive | Optimal | Intelligent | Tracking ability | Fast convergence/response | Precision | Simplicity | Disturbance rejection | Unmodeled parameter handling | Manual tuning | (Signal) noise | Chattering/energy loss |
| 1. PID | 1 | 0 | 0 | 0 | 1 | 1 | 1 | 2 | 0 | 0 | 2 | 2 | 0 |
| 2. Intelligent PID | 1 | 0 | 0 | 2 | 1 | 1 | 1 | 1 | 0 | 0 | 0 | 1 | 0 |
| 3. LQR | 0 | 2 | 1 | 0 | 1 | 1 | 0 | 1 | 1 | 0 | 1 | 1 | 0 |
| 4. LQG | 0 | 2 | 2 | 0 | 1 | 1 | 0 | 0 | 2 | 0 | 1 | 0 | 0 |
| 5. $L_1$ | 0 | 2 | 2 | 0 | 1 | 2 | 2 | 0 | 1 | 0 | 0 | 0 | 0 |
| 6. $H_1$ | 2 | 1 | 2 | 0 | 2 | 0 | 1 | 0 | 1 | 1 | 0 | 0 | 0 |
| 7. SMC | 1 | 2 | 1 | 0 | 2 | 2 | 2 | 1 | 2 | 1 | 0 | 0 | 2 |
| 8. FBL | 1 | 1 | 0 | 0 | 2 | 2 | 2 | 1 | 1 | 1 | 0 | 1 | 0 |
| 9. Backstepping | 0 | 2 | 0 | 0 | 2 | 0 | 1 | 0 | 2 | 1 | 0 | 0 | 0 |
| 10. Fuzzy logic | 1 | 1 | 1 | 2 | 1 | 1 | 1 | 1 | 1 | 0 | 1 | 0 | 0 |
| 11. Neural networks | 1 | 2 | 2 | 2 | 1 | 1 | 1 | 0 | 1 | 1 | 0 | 0 | 0 |
| 12. Genetic | 1 | 2 | 2 | 2 | 1 | 1 | 1 | 0 | 1 | 2 | 0 | 0 | 0 |

Legend: 0—low to none; 1—average; 2—high. Also, 1 through 5 (Linear); 6 through 12 (Nonlinear).

This paper has reviewed several common control algorithms that have been used on quadrotors in literature. As evident from the review, no single algorithm presents the best of the required features. It also been discussed that getting the best performance usually requires hybrid control schemes that have the best combination of robustness, adaptability, optimality, simplicity, tracking ability, fast response and disturbance rejection among other factors. However, such hybrid systems do not guarantee good performance; hence a compromise needs to be found for any control application on which of the factors would be most appropriate. The jury is still out as to what mathematical model would give the best overall performance.

The designed quadrotor will be used for game counting at the National Parks; hence it should have the following important characteristics: high endurance, low noise, high agility, high cruising and VTOL ability. The quadrotor will be a great aid for nature conservationists that are looking for innovative ways to combat poaching and get accurate animal statistics. Future possibilities with this type of quadrotor would include the possibility of equipping it with a manipulator that can do other things such as spraying of medicines or tagging.